\documentclass[prb,twocolumn,showpacs,superscriptaddress]{revtex4}

\usepackage[english]{babel}
\usepackage[applemac]{inputenc}
\usepackage[T1]{fontenc}
\usepackage{ae}
\usepackage{graphicx} 
\usepackage{amsfonts} 
\usepackage{amssymb} 
\usepackage{amsmath} 
\usepackage{latexsym} 
\usepackage{psfrag} 
\usepackage{float} 
\usepackage{color} 
\usepackage{subfigure} 
\usepackage{natbib}

\newcommand{\beq}{\begin{equation}}
\newcommand{\eeq}{\end{equation}}

\def\bra#1{\mathinner{\langle{#1}|}} 
\def\ket#1{\mathinner{|{#1}\rangle}}

\newcommand{\de}{\mathrm{d}}										

\newcommand{\desude}[1]{\frac{\de }{\de #1}}

\begin{document}

\title{Nonadiabatic stationary behaviour in a driven low-dimensional gapped system}

\author{Anna Maraga}
\affiliation{SISSA, International School for Advanced Studies, via Bonomea 265, 34136 Trieste, Italy}
\author{Pietro Smacchia}
\affiliation{SISSA, International School for Advanced Studies, via Bonomea 265, 34136 Trieste, Italy}
\author{Michele Fabrizio}
\affiliation{SISSA, International School for Advanced Studies, via Bonomea 265, 34136 Trieste, Italy}
\author{Alessandro Silva}
\affiliation{SISSA, International School for Advanced Studies, via Bonomea 265, 34136 Trieste, Italy}
\affiliation{Abdus Salam ICTP, Strada Costiera 11, 34100 Trieste, Italy}

\begin{abstract}
We discuss the emergence of nonadiabatic behavior in the dynamics of the order parameter in a low-dimensional quantum many-body system subject to a linear ramp of one of its parameters. While performing a ramp within a gapped phase seems to be the most favorable situation for adiabaticity, we show that such a change leads eventually to the disruption of the order, no matter how slowly the ramp is performed. We  show this in detail by studying the dynamics of the one-dimensional quantum Ising model subject to linear variation of the transverse magnetic field within the ferromagnetic phase, 
and then propose a general argument applicable to other systems.
\end{abstract}
\maketitle

The nonequilibrium dynamics of  isolated quantum many-body systems is one of the most active and interdisciplinary fields that emerged recently \cite{Polkovnikov2011,Dziarmaga2010,Lamacraft2012}. 
Indeed, while interest in this area has been  spurred by the opportunity to directly access the nonequilibrium dynamics in cold atom gases loaded in optical lattices \cite{bloch_review}, many of the questions addressed in that context turn out to be of importance in others, such as high energy physics \cite{Chesler2009} and cosmology \cite{Berges2004,Das2012}. In all intriguing issues addressed in the recent literature, such as the meaning and occurrence of thermalization in isolated quantum systems, or the quest for ``universal'' behavior out of equilibrium, a recurring theme has been the characterization of the 
response of a many-body system to the variation of the Hamiltonian parameters. In particular, the main focus has been on the two extremes of instantaneous changes (quenches) and slow ones (known under the oxymoron ``slow quenches''). The latter has been mostly  studied for systems driven across a quantum critical point, where a generalization of the classical Kibble-Zurek theory led to the prediction of a universal scaling of the excitation density with the speed at which the critical point is crossed  \cite{ap_adiabatic,zurek_05}, successively extended also to quenches within gapless phases \cite{degrandi_2009,eckstein_10}, where even full violation of adiabaticity may occur \cite{polkovnikov_gritsev_08}.
Specifically, universality is expected whenever the scaling dimension of the fidelity susceptibility \cite{gu_09} (or its generalization for non linear protocols) is negative, and extends to other quantities  besides the excitation density, such as the excess energy. We also mention that spontaneous generation of defects in the nonequilibrium dynamics has been observed experimentally in spinor condensates \cite{veng2}. 

Intuitive quantum mechanical arguments, rooted ultimately on the adiabatic theorem, suggest that the case of quenches within a gapped phase is much less interesting. Indeed, in this case the scaling dimension of the fidelity susceptibility is always positive, implying that the density of excitations and the excess energy always tends to zero with the square of the switching rate for linear ramps (generalization to generic power-law ramps is straightforward). This also suggests that other thermodynamics quantities share the same property \cite{degrandi_2009}, i.e., corrections with respect to their equilibrium value are quadratic in the rate \cite{Polkovnikov2011}. However, intuition indicates a different scenario when considering the order parameter in a phase with spontaneous symmetry breaking.  Since even when performing a variation of the Hamiltonian within a gapped phase an extensive amount of energy is injected, one expects to be in a situation similar to the case of finite temperature. In certain instances, for example in low-dimensional systems, the effect of temperature is the complete disruption of long-range order \cite{sachdev_97}, an effect which is very far from being a small correction.  

This work addresses this apparent contradiction by studying the dynamics of the order parameter~$m^x(t)$ in a one-dimensional quantum Ising chain after a linear variation in time of the transverse field within the ferromagnetic, ordered phase. In particular, we focus on the asymptotic value of the order parameter~$m^x(t\to\infty)$ as a function of the duration~$\tau$ of the linear ramp. 
We show that, even though the bigger~$\tau$ is the 
closer~$m^x(\tau)$ gets to its ground state value~$m_0^x$, 
nevertheless, however small~$|m^x(\tau)-m_0^x|$ is ---actually it is proportional to~$1/\tau$--- it is enough to completely disrupt the order exponentially fast in the subsequent time evolution,~$m^x(t\to\infty)\to 0$. In particular, in the stationary state the inverse correlation length turns out to depend quadratically on the ramp rate for large~$\tau$. These quadratic corrections persist also in the limit of small~$\tau$, where the reference value is that of the sudden limit~$\tau=0$. For protocols of intermediate durations in turn the inverse correlation length displays an oscillatory behavior. These results show that in low-dimensional many-body systems  an apparently small correction to adiabaticity can lead to major consequences for certain observables, even in a gapped phase.

Let us start our analysis by introducing the Hamiltonian of model
\beq
\mathcal{H}(t)= - \sum_{j=1}^L \Big[ \sigma_j^x \sigma_{j+1}^x + g(t)\, \big(\sigma_j^z -1\big)\Big],
\label{eq:hamiltonian}
\eeq
where $\sigma_j^\alpha$ are the Pauli matrices, satisfying periodic boundary conditions $\sigma_{j+L}^\alpha=\sigma^\alpha_j$, and $g(t)$ represents a linear ramp of the transverse field, i.e., $g(t)=g_0$ for~$t \leq 0$, $g(t)=g_0+(g_1-g_0)t/\tau$ for~$0<t<\tau$, and $g(t)=g_1$ for~$t \geq \tau$.  We will assume that the system is initially prepared in its ground state for $g=g_0$.

At zero temperature this model exhibits a quantum phase transition at $g_c=1$ separating a ferromagnetic phase ($g<g_c$) from a paramagnetic one ($g>g_c$), both characterized by a finite gap. At any finite temperature the system is instead paramagnetic \cite{Sachdev_book}. The order parameter  is the spontaneous magnetization along the $x$ axis, defined as $m_x=\frac{1}{L}\sum_j \langle \sigma_j^x \rangle$, which is finite in the ferromagnetic phase and zero in the paramagnetic one. As stated above, we are interested in the dynamics within the ordered phase, so we take both $g_{0} < 1$ and $g_{1} < 1$.

Performing a Jordan-Wigner transformation \cite{jordan_wigner},
\begin{subequations}
\label{eq:jordan_wigner}
\begin{align} 
\sigma_j^x &= \prod_{m<j} \left( 1 - 2 c_m^{\dagger} c_m \right)  \left( c_j + c_j^{\dagger} \right) \label{JW_trans_a}, \\
\sigma_j^z &= 1- 2 c_j^{\dagger} c_j \label{JW_trans_b},
\end{align}
\end{subequations}
with $\{c_j,c^\dagger_l\}=\delta_{jl}$ and $\{c_j,c_l\}=0$, the Hamiltonian (\ref{eq:hamiltonian}) can be written as \cite{dziarmaga_05}
\beq
\mathcal{H}(t) = P^+ \mathcal{H}^+(t) P^+ + P^- \mathcal{H}^-(t) P^-,
\eeq
where
\beq
P^{\pm}=\frac{1}{2} \left[1\pm \prod_{j=1}^L \sigma_j^z \right]
\eeq
are the projectors in the subspace with an even ($+$) or odd ($-$) number of fermions and
\begin{align} 
\mathcal{H}^{\pm}(t)= & - \sum_{i=1}^L 
\Big(c_i^{\dagger}c_{i+1} + c_i^{\dagger}c_{i+1}^\dagger + h.c.\Big) - 2g(t) c_i^{\dagger} c_{i} ,
\label{eq:fermion_hamiltonian}
\end{align} 
with the $c_i$'s obeying antiperiodic boundary conditions~$c_{L+1}=-c_1$ in the even sector and periodic boundary conditions $c_{L+1}=c_1$ in the odd one.

For finite chains the ground state is always in the even sector and the order parameter $\sigma^x_j$, which 
changes the parity of the fermion number, is strictly zero.  However, the energy gap between the lowest energy states within each sector, $\ket{\Omega_+}$ and $\ket{\Omega_-}$, vanishes exponentially 
in the thermodynamic limit and in the ferromagnetic phase, manifestation of spontaneous breaking of 
the $\mathbb{Z}_2$ symmetry. One can nonetheless recognize spontaneous symmetry breaking 
even within each separate sector through the long-distance behavior of the correlation function $R^x_r = 
\langle \Omega_{\pm}\!\mid\sigma^x_j\,\sigma^x_{j+r}\mid \! \Omega_{\pm}\rangle$, which is independent of $j$.  
Indeed, in the ferromagnetic phase, $\lim_{r\to\infty} R^x_r = m_x^2 >0$, signaling the established long-range order. We shall thence focus on the even sector, where the finite-size ground state lies, and study 
the time evolution of 
\beq
R^x_r(t) = \lim_{L \rightarrow \infty} \langle \psi_+(t)\!\mid \sigma^x_j\,\sigma^x_{j+r}\mid \!\psi_+(t)\rangle,
\label{eq:even_observables}
\eeq
where $\ket{\psi_+(t)}=\mathcal{U}(t) \ket{\Omega_+}$, being $\mathcal{U}(t)$ the evolution operator, 
 and $\ket{\Omega_+}$ the initial state assumed to be the ground state at $g=g_0$.

The Hamiltonian (\ref{eq:fermion_hamiltonian}) can be instantaneously diagonalized performing a Fourier transform $c_j=\frac{e^{i \pi/4}}{\sqrt{L}} \sum_k e^{i k j} \hat{c}_k$, with $k$ odd multiple of $\pi/L$ so 
to implement the antiperiodic boundary conditions in the even sector, followed by a Bogoliubov transformation,
\begin{align} \label{B_trans}
\begin{pmatrix}%
				\hat{c}_k	   \\
				\hat{c}_{-k}^{\dagger} \\
\end{pmatrix} = 
\begin{pmatrix}%
				u_k(t)	&- v_k(t)   \\
				v_k(t)	&u_k(t) \\
\end{pmatrix}
\begin{pmatrix}%
				\gamma_k^t \\
				{\gamma_{-k}^t }^{\dagger}	 \\
\end{pmatrix},
\end{align}
with coefficients $u_k(t)=\frac{1}{\sqrt{2}} \sqrt{1+\frac{g(t)-\cos(k)}{\epsilon_k(t)}}$, $v_k(t)=-\frac{1}{\sqrt{2}} \sqrt{1-\frac{g(t)-\cos(k)}{\epsilon_k(t)}}$, and eigenvalues $\epsilon_k(t) = \sqrt{1+g^2(t) - 2 g(t) \cos(k)}$. 
The instantaneous ground state is $\ket{\Omega_+}_t=\prod_{k>0} \left(u_k(t)+v_k(t) \hat{c}^\dagger_{-k} \hat{c}^\dagger_k \right) \ket{0}$, with $c_k \ket{0}=0$ $\forall k$, and $\ket{0} = \bigoplus_{k>0} \ket{0}_k$. 

The dynamics induced by the linear ramp $g(t)$ can be described through the density matrix $\rho(t)=\ket{\psi_+(t)} \bra{\psi_+(t)}$ that, since the $k$ modes are mutually independent, has the form $\rho(t) =\bigotimes_{k>0} \rho_k(t)$, where~$\rho_k(t)$  in the basis $\{\ket{0}_k, \hat{c}^\dagger_{-k} \hat{c}^\dagger_k \ket{0}_k\}$ is given by
$\rho_k(t)=\frac{1}{2}(\mathbf{\hat{1}} + f_{1,k}(t)\mathbf{\hat{\tau}}^z+f_{2,k}(t) \mathbf{\hat{\tau}}^x+f_{3,k}(t) \mathbf{\hat{\tau}}^y)$,
where $\hat{\tau}^{x,y,z}$ are the Pauli matrices.
In vector notation the coefficients~$\hat{f}_k=\left( f_{1,k}, f_{2,k}, f_{3,k} \right)^T$ satisfy the simple equation
\beq
\desude{t} \hat{f}_k(t)=\mathbf{\hat{R}}_k \hat{f}_k(t),
\label{eq_motion_f}
\eeq
where the matrix $\mathbf{\hat{R}}_k=-4(g(t)-\cos(k))\hat{\mathbf{X}}_1+4\sin(k)\hat{\mathbf{X}}_2$, where $\hat{\mathbf{X}}_i$ $(i=1,2,3)$ are the $3  \times 3$ generators of rotations. Notice that the initial conditions $f_{1,k}(0)=\frac{g_0-\cos k}{\epsilon_k(0)}$, $f_{2,k}(0)=-\frac{\sin k}{\epsilon_k(0)}$, $f_{3,k}(0)=0$.

From the evolution of the density matrix we can calculate 
$R_r^x(t)$ of Eq. (\ref{eq:even_observables}). We start by writing $R_r^x(t)$ as\cite{LSM}
\beq
R_r^x(t)=\langle B^t_j A^t_{j+1} B^t_{j+1} \cdots A^t_{j+r-1} B^t_{j+r-1} A^t_{j+r}\rangle_0,
\label{eq:corr_x}
\eeq
where $A^t_j=c_j(t)+c^\dagger_j(t)$, $B^t_j=c^\dagger_j(t)-c_j(t)$, $c_j(t)=\mathcal{U}^\dagger(t) \,c_j \, \mathcal{U} (t)$, and $\langle \cdots \rangle_0$ denotes the average over the initial state. 
\begin{figure}
\includegraphics[width=.9\columnwidth]{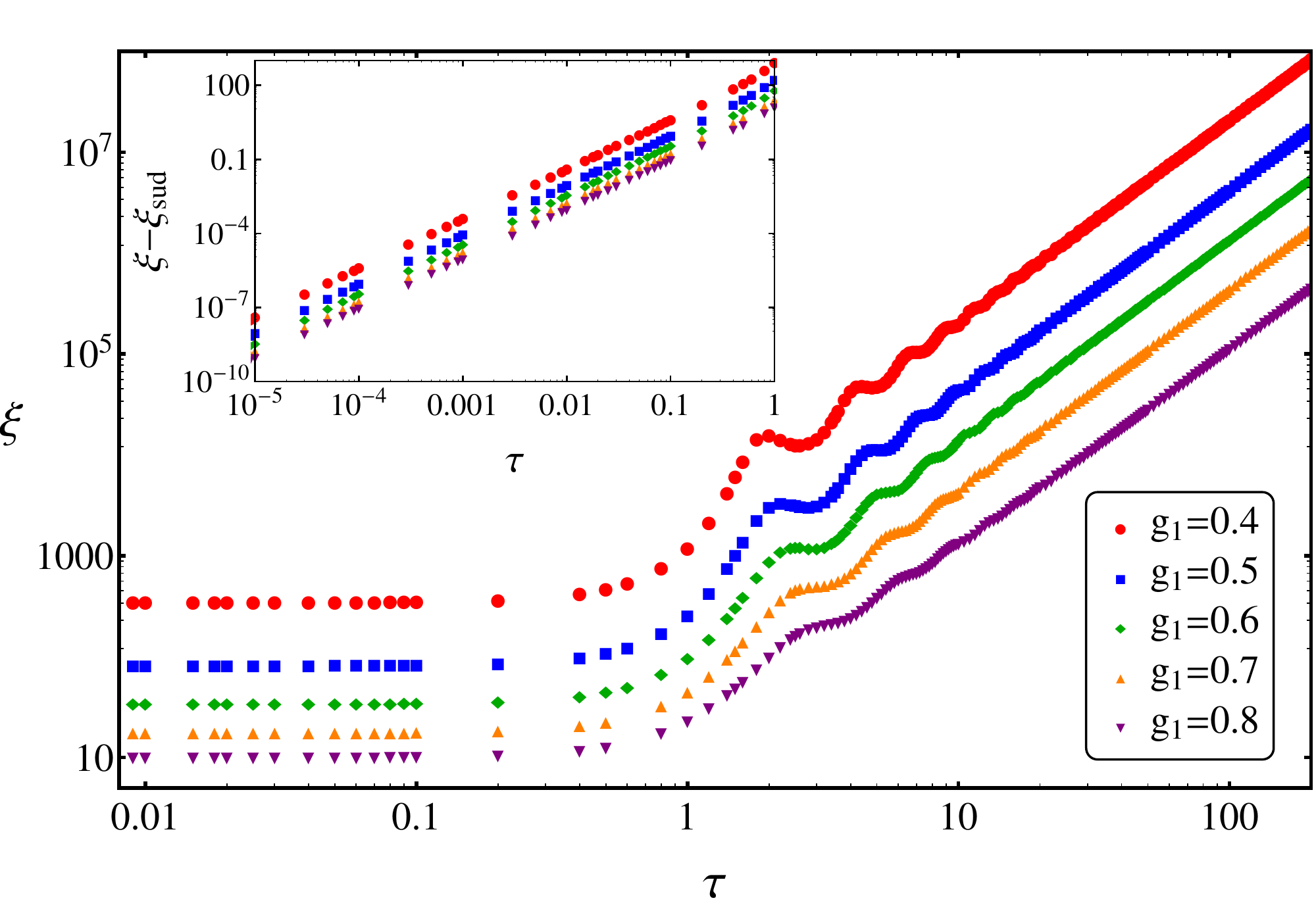}
\caption{(Color online) Log-log plot of the correlation length~$\xi$ as a function of the duration $\tau$ of the linear ramp for initial transverse field $g_0=0.3$ and different final values of $g_1$. $\xi_{sud}$ is the value of the correlation length for a sudden quench from $g_0$ to $g_1$.}
\label{fig:corr_length}
\end{figure}

\begin{figure}
\includegraphics[width=.9\columnwidth]{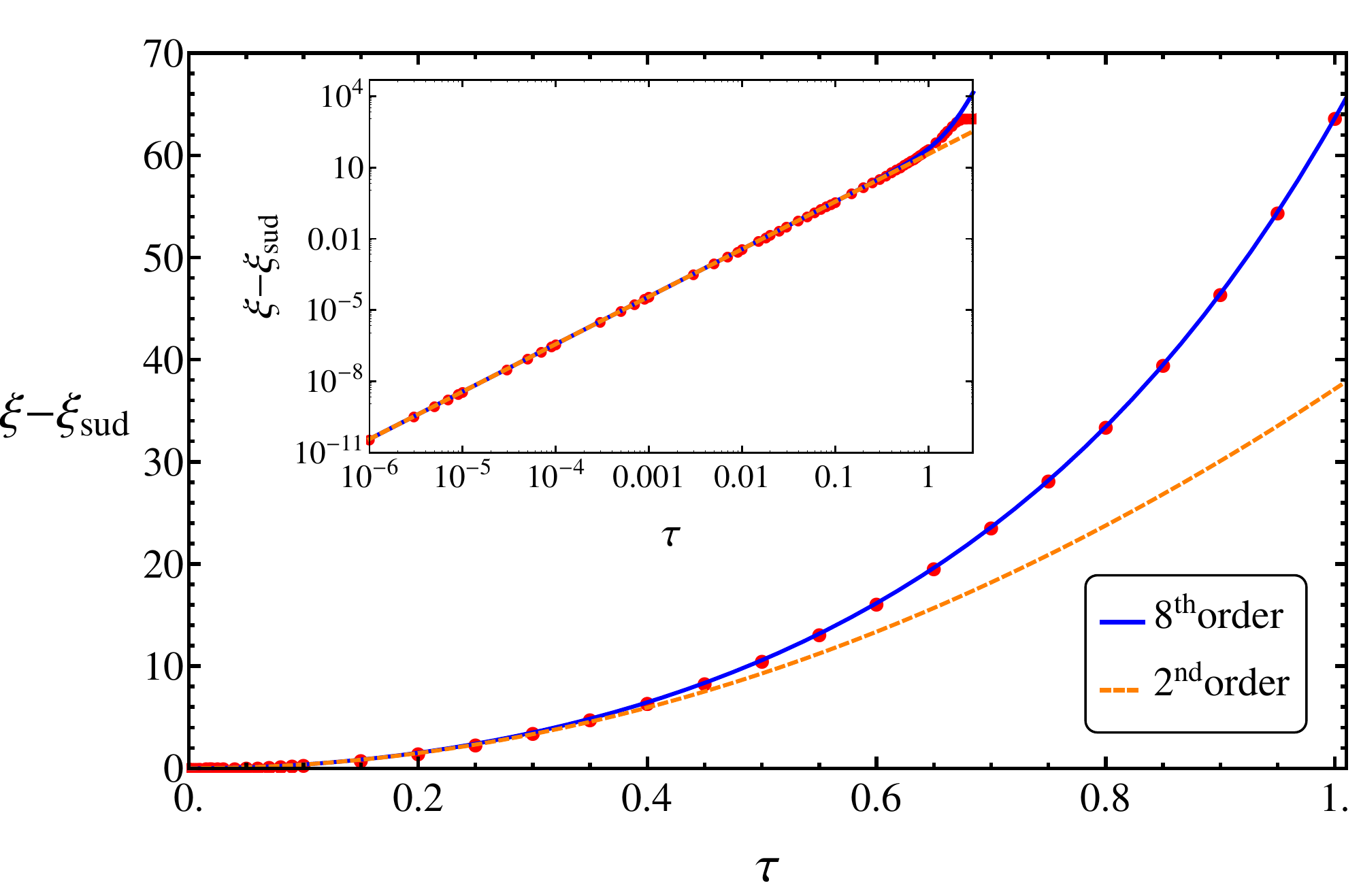}
\caption{(Color online) Correlation length~$\xi$ as a function of the duration $\tau$ for $g_0=0.3$ and $g_1=0.6$. The numerical results (red circles) are compared with the perturbative expansion up to second and eighth order. The inset shows the same plot in log-log scale.}
\label{fig:small_tau}
\end{figure}

Using Wick's theorem, Eq. (\ref{eq:corr_x}) can be expressed in terms of the contractions of the $A_j$'s and $B_j$'s, which in terms of the functions $f_{1,k}$, $f_{2,k}$, and $f_{3,k}$ read
\begin{subequations}
\beq
\langle A^t_j A^t_l \rangle_0=\delta_{jl}-\frac{1}{L} \sum_k e^{i k(j-l)} f_{3,k}(t),
\eeq
\beq
\langle B^t_j B^t_l \rangle_0=-\delta_{jl}-\frac{1}{L} \sum_k e^{i k(j-l)} f_{3,k}(t),
\eeq
\beq
\langle B^t_j A^t_l \rangle_0 =-\frac{1}{L} \sum_k e^{i k (j-l)} [f_{1,k}(t)+i f_{2,k}(t)]
\eeq

\end{subequations}
Using these equations one may easily compute the time evolution of the order parameter (see Supplemental Material). Right after the ramp, the order parameter is $m^x=m_0^x+\delta m^x$, where $m^x_0$ is the value it would have in the ground state of the final Hamiltonian, while~$\delta m^x \propto 1/\tau$ is a correction. Unlike classical systems, where these corrections would lead to a small precession of the magnetisation around its equilibrium value, in a quantum low-dimensional system this state is dynamically very fragile, and the subsequent time evolution produces a collapse of the magnetization. Let us see this considering the stationary state, that is, for~$t \rightarrow \infty$ after taking 
the thermodynamic limit, i.e., replacing discrete sums over $k$ with integrals. For $t > \tau$, $g(t)=g_1$ is constant, so we can readily integrate Eqs. (\ref{eq_motion_f}) in terms of the boundary values $f_{1,k}(\tau)$, $f_{2,k}(\tau)$, and $f_{3,k}(\tau)$. The solution consists in a stationary part plus oscillatory terms with frequency~$4 \epsilon_k(\tau)$, which vanish for $t\to\infty$ once integrated over $k$. We thus find that $\langle A_j A_l\rangle_0 \rightarrow \delta_{jl}$, $\langle B_j B_l \rangle_0 \rightarrow -\delta_{jl}$, and $\langle B_j A_l \rangle_0 \rightarrow C(j-l+1)$, with
\beq
C(r)= \int_{-\pi}^\pi \frac{ dk}{2 \pi} \frac{\cos \left(k r \right)-g_1 \cos \left[k(r-1)\right]}{1+g_1^2-2 g_1 \cos k}(1-2 n_k)
\label{eq:d_definition}
\eeq
with $n_k=\langle {\gamma^\tau}^\dagger_k \gamma_k^\tau \rangle_t$ the occupation numbers in the evolved state, which are actually time independent for~$t>\tau$ and given by  
\beq
1-2 n_k=\frac{( g_1-\cos k) f_{1,k}(\tau)-\sin k f_{2,k}(\tau)}{\sqrt{1+g_1^2-2 g_1 \cos k}}.
\label{eq:n_k}
\eeq
We note that disregarding the oscillatory terms is equivalent to stating that the stationary value, being the correlation a local observable, can be computed in the diagonal ensemble, which is completely determined by the occupation numbers $n_k$. 

\begin{figure}[t]
\includegraphics[width=.9\columnwidth]{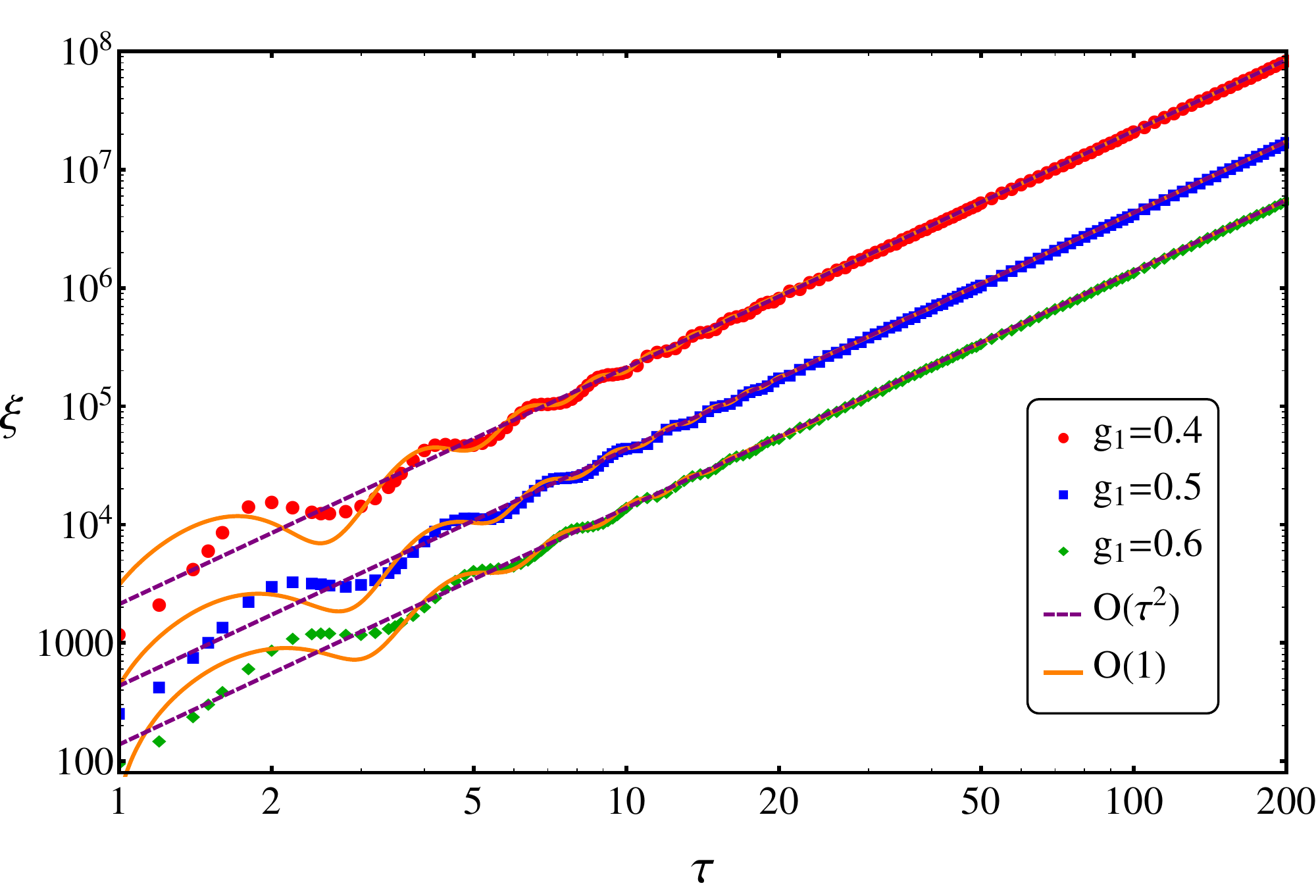}
\caption{(Color online) Log-log plot of the correlation length~$\xi$ as a function of the duration $\tau$ of the linear ramp for initial transverse field $g_0=0.3$ and different final values of $g_1$. Numerical results are compared with the predictions of adiabatic perturbation theory at two different orders.}
\label{fig:large_tau}
\end{figure}
As in equilibrium, the correlation $R_r^x$ can be expressed as a $r \times r$ Toeplitz determinant,
\beq
R_r^x =\begin{vmatrix}
C(0) & C(-1) & \dots & C(-r+1) \\
C(1) & C(0) & \dots & C(-r+2)\\
\vdots & \vdots & \ddots & \vdots\\
C(r-1) & C(r-2) & \dots & C(0)\\
\end{vmatrix},
\label{eq:det}
\eeq
whose asymptotic behavior in the limit $r \rightarrow \infty$ has to be determined. To this end ,we first note that~$C(r)=\frac{1}{2 \pi} \int_{-\pi}^\pi dk \tilde{C}(k) e^{-i k r}$, with 
\beq
\tilde{C}(k)=\left(\frac{1-g_1 e^{i k}}{1-g_1 e^{-i k}} \right)^{1/2} \left(1-2 n_k \right).
\eeq
In terms of the complex variable $z=e^{i k}$ the function~$\tilde{C}(z)$ has zero index around the unit circle and is non vanishing, as long as $n_k < 1/2$, $\forall k$, a condition that has been verified numerically and perturbatively, and is equivalent to say that the effective temperature of all the modes is less than infinity. Under this condition we can apply the strong Szeg\H{o} lemma \cite{szego}, which tells us that~$R_r^x \sim e^{-r/\xi}$, with the inverse correlation length given by
\beq
\xi^{-1}=-\frac{1}{2 \pi} \int_{-\pi}^\pi \!\!\!\! dk \ln \left(1-2 n_k \right).
\label{eq:corr_length}
\eeq
Therefore, whenever $n_k \neq 0$, the correlation length is finite, implying that $R_r^x$ goes to zero exponentially hence that the order parameter is zero. Such a condition is verified for any finite duration of the linear ramp, implying that adiabaticity is broken for the order parameter. From Eq. (\ref{eq:corr_length}) we observe that a tiny deviation of the occupation numbers with respect to their equilibrium value ($n_k=0$) translates into a comparably small inverse correlation length.
Nonetheless, such small quantitative corrections lead to a completely different behavior of the correlation function $R_r^x$ and of the order parameter.

Figure \ref{fig:corr_length} shows the correlation length as a function of~$\tau$ for different ramps computed by numerically solving Eqs.~(\ref{eq_motion_f}) and evaluating Eq. (\ref{eq:corr_length}). We can see that for long durations the correlation length grows quadratically, while for $\tau$ of order one it displays oscillations. The inset of the figure shows that also for small $\tau$ the growth of $\xi$ above the sudden-quench value is quadratic. The two limiting cases of slow and sudden quenches can be captured by two different perturbative expansions (more details can be found in the Supplemental Material).

For small $\tau$ the result of the perturbative expansion of Eqs. (\ref{eq_motion_f}) at the leading order is
\begin{widetext}
\beq
\xi(\tau)=-\frac{1}{\ln \left[\frac{1+g_0 g_1+\sqrt{(1-g_1^2)(1-g_0^2)}}{2} \right]}+\tau^2 \frac{2(g_1-g_0)^2 \left[1+g_0 g_1-\sqrt{(1-g_1^2)(1-g_0^2)}\right]}{3(g_0+g_1)^2\ln^{2} \left[\frac{1+g_0 g_1+\sqrt{(1-g_1^2)(1-g_0^2)}}{2} \right]}+O(\tau^4),
\label{eq:corr_length_small_tau}
\eeq
\end{widetext}
where the first term is the result for a sudden quench~($\xi_{sud}$). Higher order can be straightforwardly computed. In particular we notice that only even powers of $\tau$ are present in the expansion, and all computed corrections are even under $g_0 \leftrightarrow g_1$, i.e., inversion of the ramp. Figure \ref{fig:small_tau} shows a comparison between the perturbative and the numerical results, and we can see that the agreement is excellent up to $\tau \simeq 1$ provided corrections up to eighth order are taken into account.

For large $\tau$, instead, one can use the adiabatic perturbation theory described in Ref. \onlinecite{ortiz_2008}, which predicts that the occupation numbers $n_k$ for large $\tau$ vanish as $1/\tau^2$ in an oscillating fashion. This is actually the source of oscillations observed in $\xi$. Indeed, by applying the adiabatic perturbation theory one obtains
\begin{widetext}
\beq
\xi(\tau)=\frac{64 (1-g_0^2)^3 (1-g_1^2)^3}{(g_1-g_0)^2 \left[(1-g_0^2)^3+(1-g_1^2)^3\right]} \tau^2+f(\tau) \sqrt{\tau}+\Lambda+O(\tau^{-1/2}),
\label{eq:corr_length_large_tau}
\eeq
\end{widetext}
where $f(\tau)$ is an oscillating function and $\Lambda$ is a constant (see the Supplemental Material). Thus, the relative oscillations of the correlation length goes to zero as $\tau^{-3/2}$. Also in this case all the corrections are invariant under the transformation $g_0 \leftrightarrow g_1$.
Figure \ref{fig:large_tau} shows a comparison between this adiabatic perturbative expansion and the numerical data. We see that by including correction up to $O(1)$ there is quite good agreement for $\tau \gtrsim 10$.

In conclusion, we have shown that the stationary value of the order parameter of a one-dimensional quantum Ising model does not behave in an adiabatic way within the ferromagnetic phase, 
however small the switching rate of the transverse field is. This occurs in spite of the fact that the Hamiltonian is gapped, which in principle is the most favorable situation for an adiabatic evolution. Such a behavior of the order parameter has to be expected whenever the system has a phase transition only at zero temperature and it is driven within the ordered phase. Indeed a finite density of excitations~$n_{ex} \sim 1/\tau^2$ will always be generated and in this situation will be always sufficient to destroy order. From this, one can estimate also the behavior of the correlation length, which, following the same reasoning as the Kibble-Zurek argument, will be $\xi \sim 1/n_{ex}^{1/d} \sim \tau^{2/d}$, with $d$ being the dimension of the system. A natural question that comes up is what happens instead in an analogous system where the transition survives at finite temperature. One possibility is that there is a transition in the value of the order parameter as a function of $\tau$, namely, for sufficiently slow ramp its asymptotic value is expected to be finite, while it should go to zero for fast ramps. If this is really the case, and in the affirmative case if the value of the order parameter is vanishing or not are interesting questions to consider in following studies.

{\it Acknowledgment -}  We acknowledge support by the European Union, Seventh Framework Programme, under the project GO FAST, Grant Agreement No. 280555.

\bibliography{Nonequilibrium}

\end{document}